\documentstyle{camera}

\newcommand{\AmS}{{\protect\the\textfont2
  A\kern-.1667em\lower.5ex\hbox{M}\kern-.125emS}}

\newcommand{\beq}{\begin{equation}}
\newcommand{\eeq}{\end{equation}}
\newcommand{\bea}{\begin{eqnarray}}
\newcommand{\eea}{\end{eqnarray}}

\def\dm2{\Delta m^2}
\def\sq2{sin^2(2\Theta)}

\input epsf

\begin{document}

%
\title{ARGO-YBJ HIGHLIGHTS}

%
\author{GIUSEPPE DI SCIASCIO \\ on behalf of the ARGO-YBJ collaboration}

%
\organization{INFN, Sez. Roma Tor Vergata\\ Via della Ricerca
Scientifica 1, I-00133 Roma, Italy}

\maketitle

\begin{abstract}
The ARGO-YBJ experiment is a multipurpose detector exploiting the full-coverage approach at very high altitude. The apparatus is in stable data taking since November 2007 at the YangBaJing Cosmic Ray Laboratory (Tibet, P.R. China, 4300 m a.s.l., 606 g/cm$^2$). In this paper we report the main results in Gamma-Ray Astronomy and Cosmic Ray Physics after about 3 years of operations.
\end{abstract}
\vspace{1.0cm}

\section{The detector}

The ARGO-YBJ experiment, located at the YangBaJing Cosmic Ray
Laboratory (Tibet, P.R. China, 4300 m a.s.l., 606 g/cm$^2$), is the only air shower array exploiting the full-coverage approach at very high altitude currently at work. ARGO-YBJ is a multipurpose detector which faces a wide range of fundamental issues in Cosmic Ray and Astroparticle Physics with a low energy threshold (a few hundred GeV).

The apparatus is constituted by a central carpet $\sim$74$\times$
78 m$^2$, made of a single layer of Resistive Plate Chambers
(RPCs) with $\sim$93$\%$ of active area, enclosed by a guard ring
partially instrumented ($\sim$20$\%$) up to $\sim$100$\times$110
m$^2$. The apparatus has a modular structure, the basic data
acquisition element being a cluster (5.7$\times$7.6 m$^2$),
made of 12 RPCs (2.8$\times$1.25 m$^2$ each). Each chamber is
read by 80 external strips of 6.75$\times$61.8 cm$^2$ (the spatial pixels),
logically organized in 10 independent pads of 55.6$\times$61.8
cm$^2$ which represent the time pixels of the detector. 
The read-out of 18360 pads and 146880 strips are the experimental output of the detector (Aielli G. et al., 2006). The RPCs are operated in streamer mode by using a gas mixture (Ar 15\%, Isobutane 10\%, TetraFluoroEthane 75\%) for high altitude operation. The high voltage settled at 7.2 kV ensures an overall efficiency of about 96\% (Aielli G. et al., 2009a).
The full detector is composed of 153 clusters for a total active surface of $\sim$6700 m$^2$. 
The detector is connected to two different data acquisition
systems, working independently, and corresponding to the two
operation modes, shower and scaler. In shower mode, for each event
the location and timing of every detected particle is recorded,
allowing the reconstruction of the lateral distribution and the arrival direction (Di Sciascio G. et al., 2007). In scaler mode the
total counts on each cluster are measured every 0.5 s, with limited information
on both the space distribution and arrival direction of the
detected particles, in order to lower the energy threshold down to $\sim$1 GeV (Aielli G. et al., 2008, 2009b, 2009c).

Since November 2007 the detector is in stable data taking
with a multiplicity trigger N$_{pad}\geq$20 and a duty cycle $\geq
85\%$: the trigger rate is about 3.6 kHz. 
In this paper we report the main results in Gamma-Ray Astronomy and Cosmic Ray Physics after about 3 years of operations.

\begin{figure}[t!]
\begin{minipage}[t]{.5\linewidth}
\begin{center}
\vspace{0.1cm}
\epsfysize=5.5cm 
\epsfbox{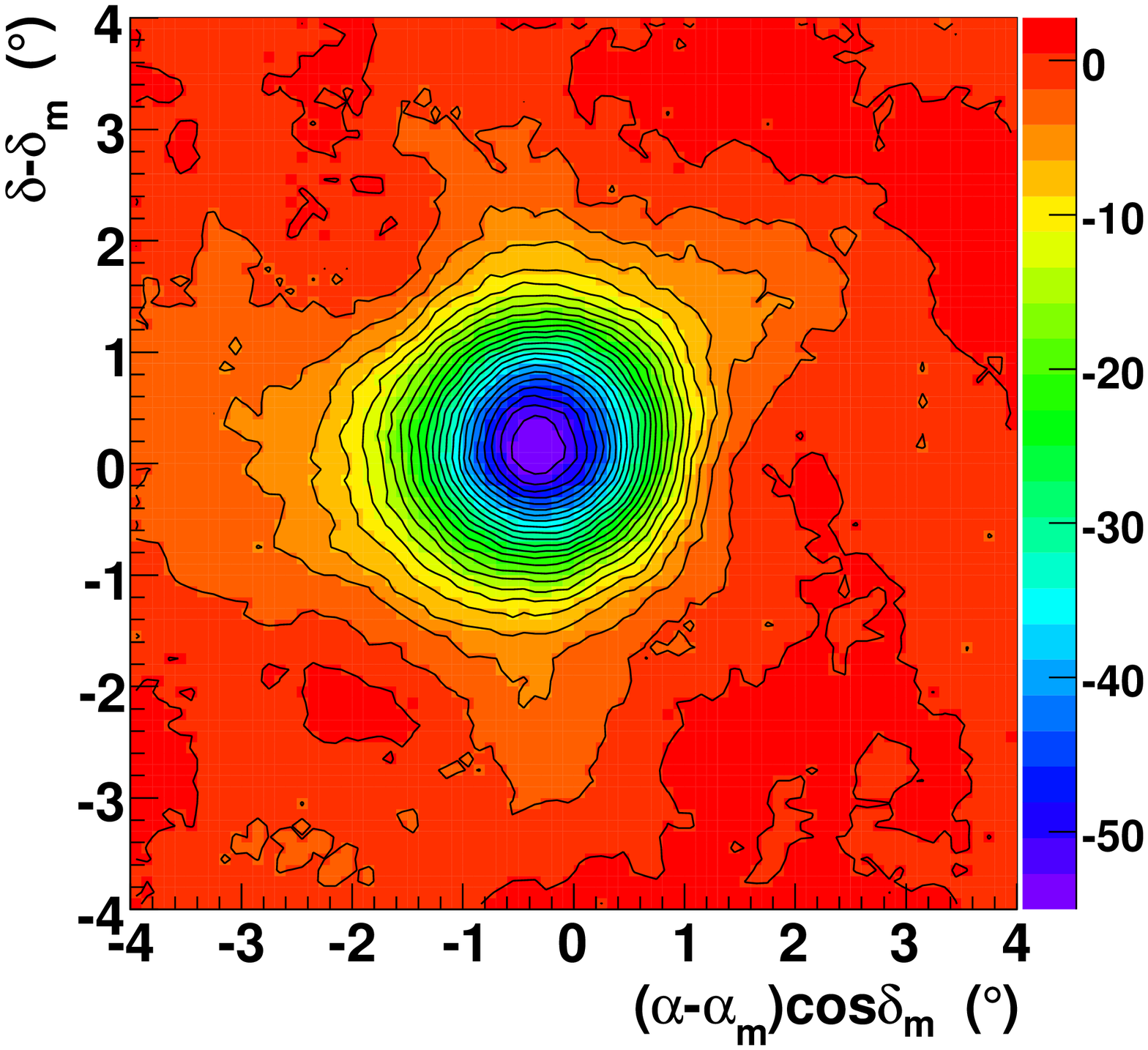}
  \end{center}
\end{minipage}\hfill
\begin{minipage}[t]{.47\linewidth}
  \begin{center}
\vspace{0.9cm}
\epsfysize=4.cm 
\hspace{0.5cm}
\epsfbox{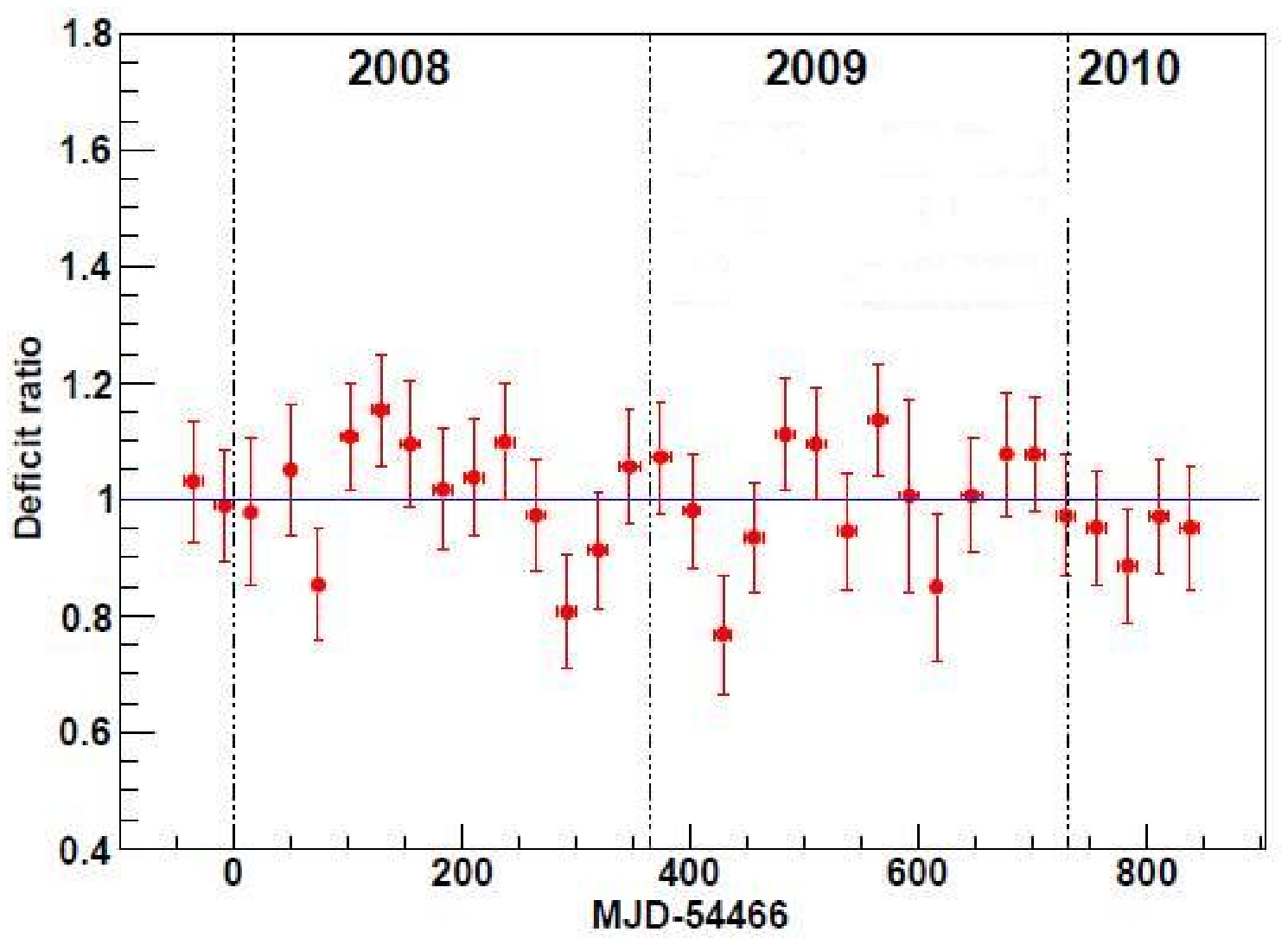}
  \end{center}
\end{minipage}\hfill
\vspace{-0.4cm} \caption[h]{Left plot: Moon shadow significance map for events with N$\geq$100. The color scale gives the significance. Right plot: the ratio of the observed deficit counts to the expected ones as a function of the observation time.} \label{fig:moon1}
\end{figure}
%

\subsection{Detector performance}

The performance of the detector and the operation stability are continuously monitored by observing the Moon shadow, i.e., the deficit of cosmic rays (CR) detected in its direction. 
Indeed, the size of the deficit allows the measurement of the angular
resolution and its position allows the evaluation
of the absolute pointing accuracy of the detector. In addition,
positively charged particles are deflected towards East due to the
geomagnetic field by an angle $\sim 1.6^{\circ}Z/E[TeV]$. Therefore, the observation of the displacement of the Moon provides a direct calibration of the relation between shower size and primary energy.

On the left side of Fig.\ref{fig:moon1} a significance map of the Moon region is shown. With all data from July 2006 to December 2009 (about 3200 hours on-source in total) we observed the CR Moon shadowing effect with a significance of about 55 standard deviations (s.d.).
The measured angular resolution is better than 0.5$^{\circ}$ for showers with energies E $>$ 5 TeV and the overall absolute pointing accuracy is at a level of $\sim$0.1$^{\circ}$.
The energy scale error of ARGO-YBJ is estimated to be less than 18\% in the energy range 1 - 30 TeV/Z.
On the right side of Fig.\ref{fig:moon1} the ratio of the observed deficit count to the expected one evaluated on a monthly basis is shown as a function of the observation time for events with N$>$100. The amount of CR deficit provides a good measure of the size of the shadow, therefore of the angular resolution.
The resulting stability of the event deficit is at a level of 10\% in the period November 2007 - March 2010.

\section{Gamma-Ray Astronomy}

\begin{figure}[t!]
\begin{minipage}[t]{.47\linewidth}
\begin{center}
\epsfysize=4.5cm \hspace{0.5cm}
\epsfbox{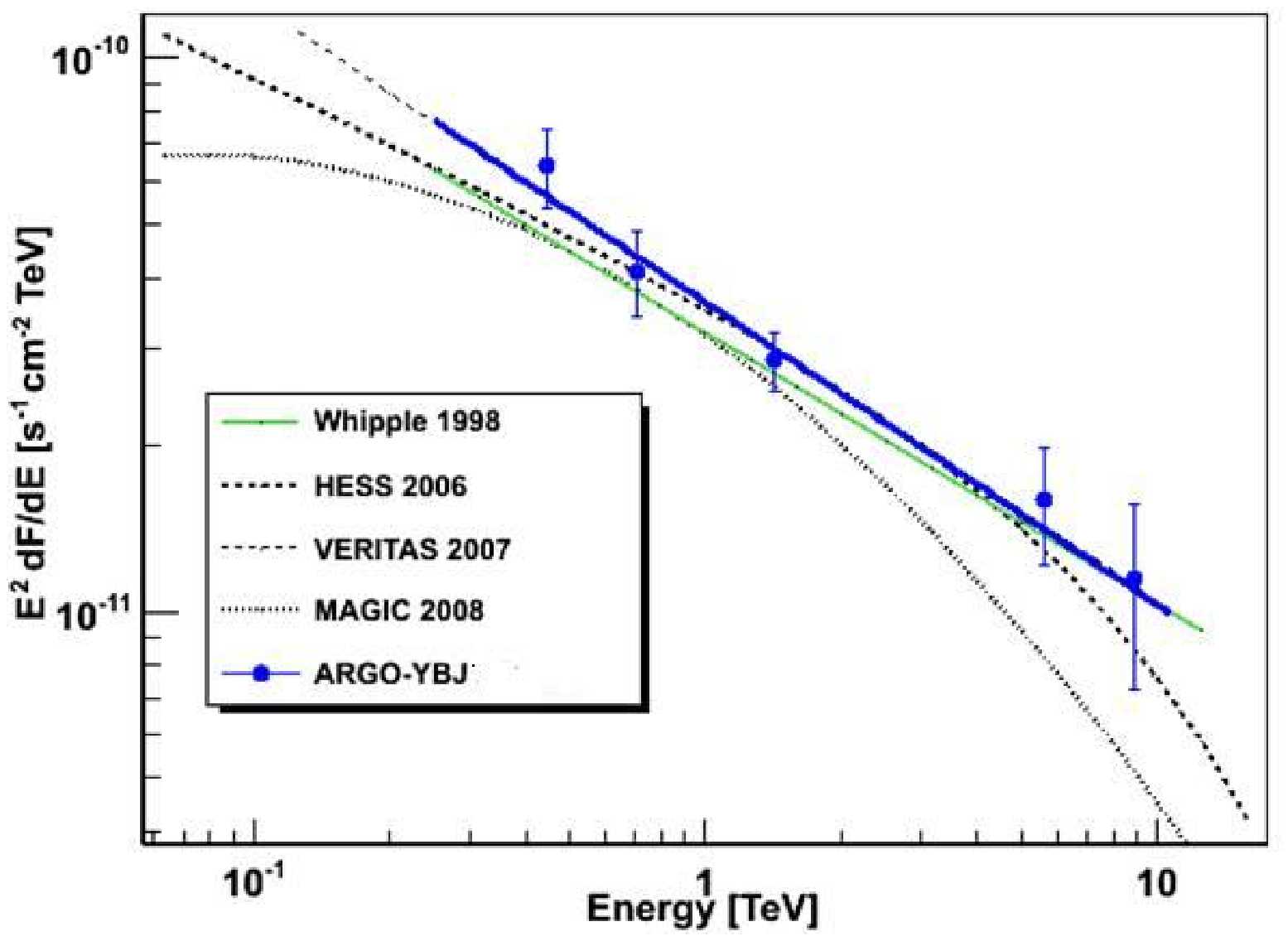} \vspace{-0.5cm}
\caption[h]{The Crab Nebula energy spectrum measured by ARGO-YBJ compared with the results of some other detectors.} 
\label{fig:crab08}
  \end{center}
\end{minipage}\hfill
\begin{minipage}[t]{.47\linewidth}
  \begin{center}
\epsfysize=4.5cm \hspace{0.5cm}
\epsfbox{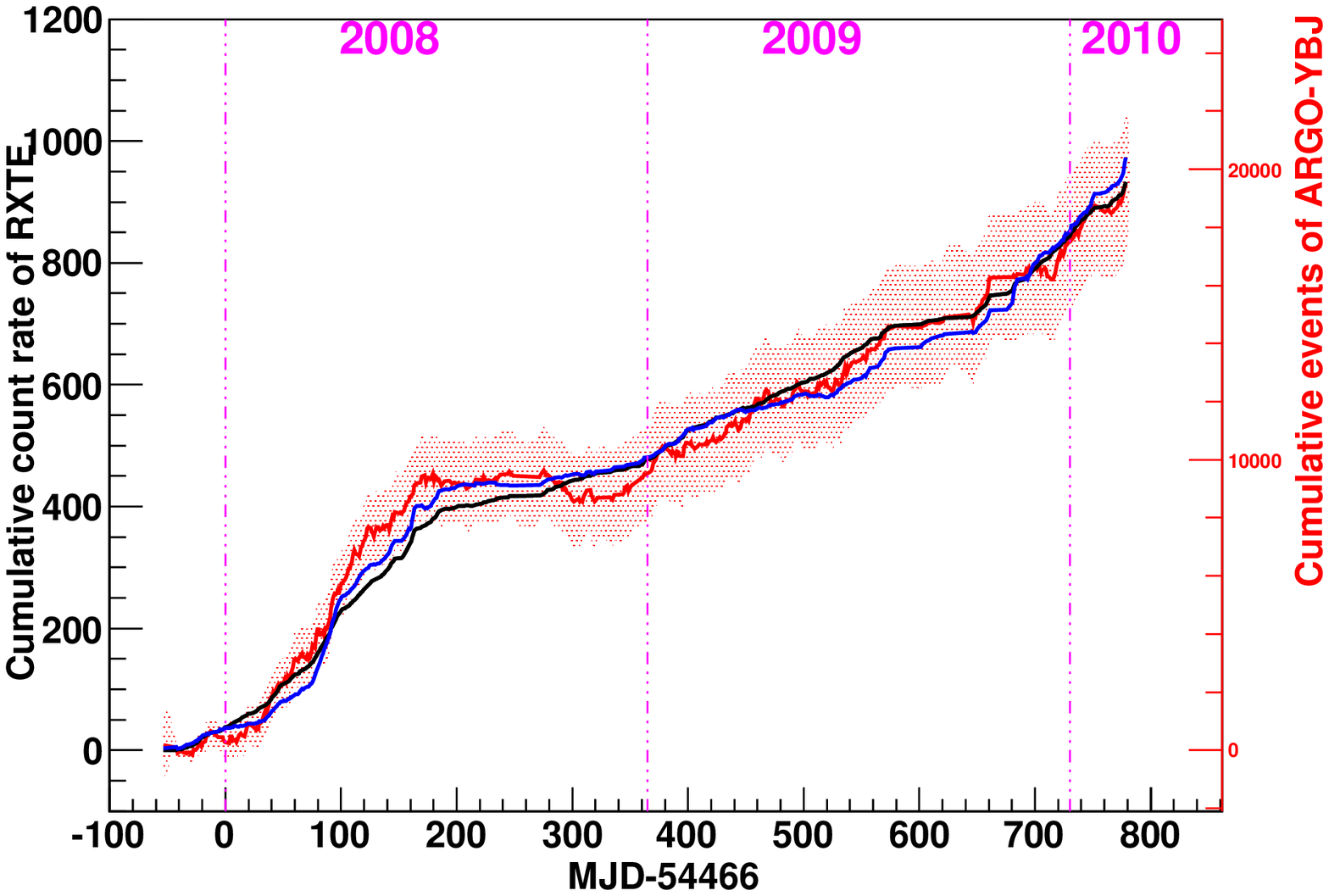} \vspace{-0.5cm}
\caption[h]{Cumulative light curve from Mrk421 measured by ARGO-YBJ compared with RXTE/ASM (black curve) and Swfit (blue curve) X-ray data.} 
\label{fig:mrk421_cumul}
  \end{center}
\end{minipage}\hfill
\end{figure}
%

With about 2 years of data we observed 3 sources with a significance greater than 5 s.d.: Crab Nebula, Mrk421 and MGRO J1908+06.

We observed the Crab signal with a significance of 14 s.d. in about 800 days at a median energy of $\sim$2 TeV, without any event selection or gamma/hadron discrimination algorithm. 
In Fig.\ref{fig:crab08} the energy spectrum of the Crab Nebula measured by the ARGO-YBJ experiment is compared to other observations. The data can be parameterized with the formula: dN/dE = (4.1$\pm$0.6)$\cdot$10$^{-11}\cdot$ (E/1 TeV)$^{-2.7\pm 0.2}$ cm$^{-2}$ s$^{-1}$ TeV$^{-1}$. Our result is in good agreement with other experiments.

Mrk421 was the first source detected by the ARGO-YBJ experiment in July 2006 when the detector started recording data with only the central carpet and was in commissioning phase.
In the last 2 years we observed the blazar Mrk421 with a total significance of about 12 s.d., averaging over quiet and active periods. Indeed, as it is well known, this AGN is characterized by a strong flaring activity both in X-rays and in TeV $\gamma$-rays. ARGO-YBJ detected different TeV flares in correlation with X-ray observations, as can be seen from Fig. \ref{fig:mrk421_cumul} where the ARGO-YBJ cumulative events per day is compared with the cumulative events per second of the RXTE/ASM and Swift satellites.
To get simultaneous data, 556 days have been selected for this analysis.
The X-ray/TeV correlation is quite evident over more than 2 years. The steepness of the curve gives the flux variation, therefore we observed an active period at the beginning of 2008 followed by a quiet phase. 

\begin{figure}[t!]
\begin{minipage}[t]{.47\linewidth}
\begin{center}
\epsfysize=4.8cm \hspace{0.5cm}
\epsfbox{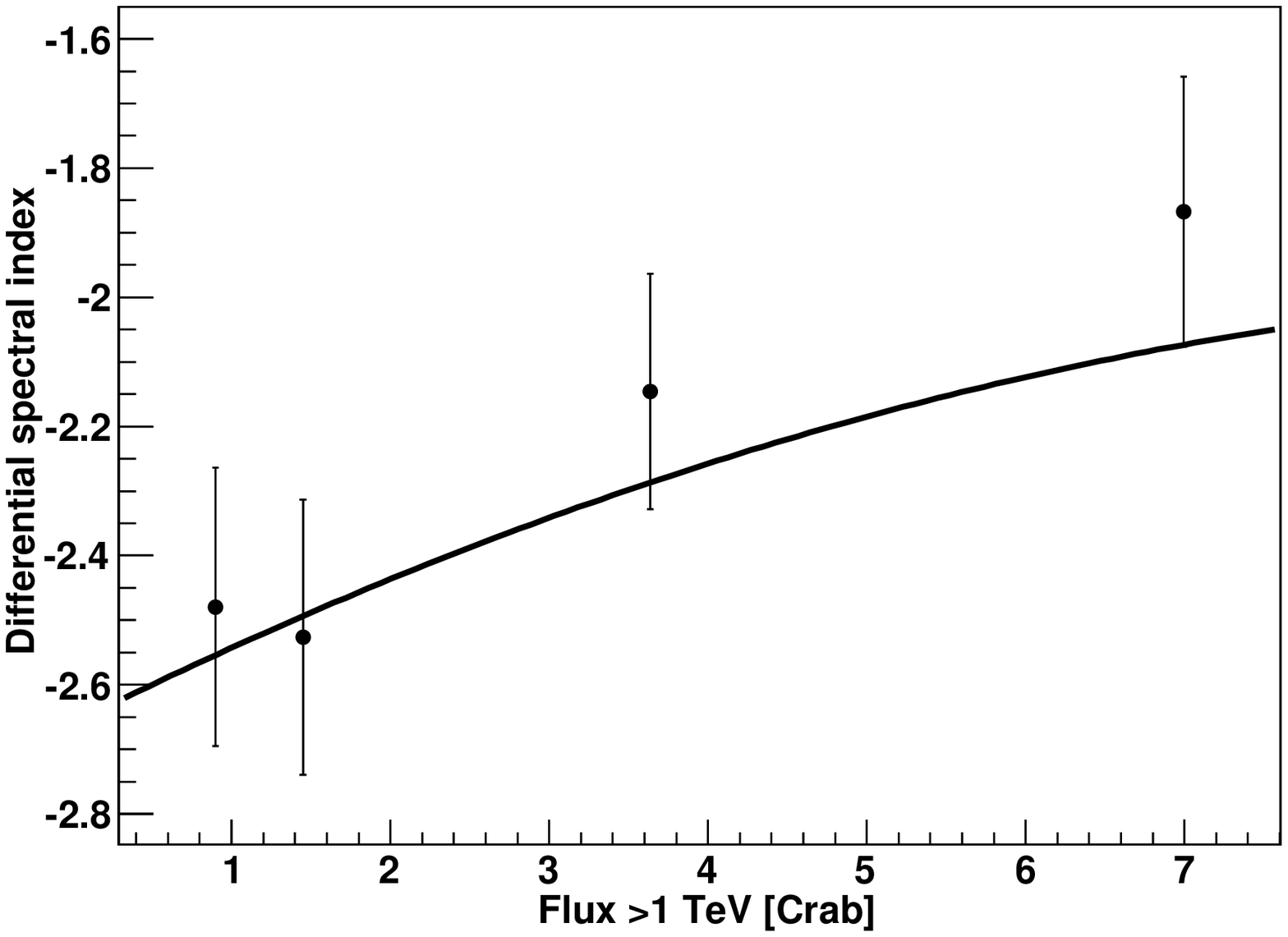} \vspace{-0.5cm}
\caption[h]{Mrk421 spectral index vs. flux above 1 TeV in Crab units. The solid line is the function obtained in Krennrich F. et al. (2002).} 
\label{fig:mrk421-spindex}
  \end{center}
\end{minipage}\hfill
\begin{minipage}[t]{.47\linewidth}
  \begin{center}
\epsfysize=4.8cm \hspace{0.5cm}
\epsfbox{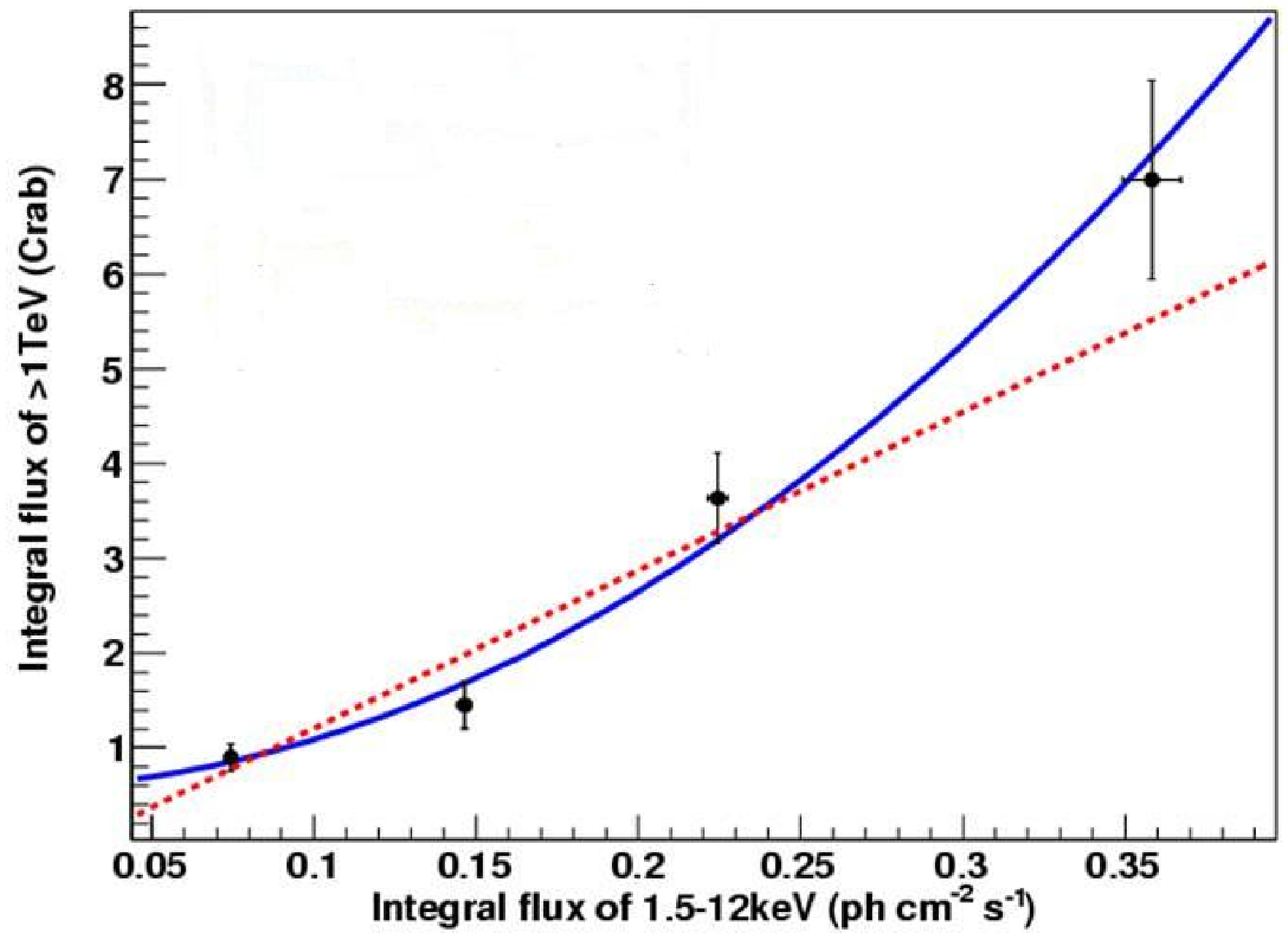} \vspace{-0.5cm}
\caption[h]{ARGO-YBJ $>$1 TeV flux vs. RXTE/ASM X-ray flux. 
Solid line: quadratic best-fit ($\chi^2$/dof= 1.9/2). Dotted line: linear best-fit ($\chi^2$/dof= 7.7/2).} 
\label{fig:mrk421-quadratic}
  \end{center}
\end{minipage}\hfill
\end{figure}
%
Cross-correlation functions were measured to quantify the degree of correlation and the phase difference (time lag) between the variations in X-ray and TeV bands, using the Discrete Correlation Function (DCF) of Edelson \& Krolik (1998). The DCF gives the linear correlation coefficient for two light curves as a function of the time lag between them.
For X-rays we consider two different bands (2-12 and 9-15 keV). 
The distributions peak around zero and the correlation coefficients around zero are $\sim$0.78. The peak positions, evaluated by fitting the distributions with a Gaussian function around the maximum, are -0.14$_{-0.85}^{+0.86}$ and -0.94$_{-1.07}^{+1.05}$ days for RXTE/ASM and Swift data, respectively.
Therefore, no significant time lag is found between X-ray data and TeV ARGO-YBJ observations.

In order to study the Spectral Energy Distribution (SED) of Mrk421 at different flux levels, both X-ray and TeV data have been divided into 4 bands based on the RXTE/ASM counting rate: 0-2, 2-3, 3-5, $>$5 cm$^{-2}$s$^{-1}$. For each band a SED with the average flux is calculated for different X-ray and $\gamma$-ray energies.
The RXTE/ASM detector is monitoring Mrk421 in three energy bands: 1.5-3, 3-5 and 5-12 keV. By assuming a power-law X-ray energy spectrum we calculated the spectral indices in the 4 flux bands: -2.02$\pm$0.08, -2.05$\pm$0.03, -2.15$\pm$0.03 and -2.43$\pm$0.04. A correlation is clearly recognized in the sense that higher fluxes are accompanied by harder energy spectra. This result, obtained by averaging the Mrk421 emission over about 2 years, extends similar conclusions obtained by averaging observations over much shorter periods (Rebillot P.F. et al., 2006).

With this information we studied the relation between TeV spectral index and flux. The TeV flux measured by ARGO-YBJ in the different bands ranges from 0.9 to about 7 Crab units.
As shown in Fig. \ref{fig:mrk421-spindex} the TeV spectral index hardens with increasing flux in agreement with the results obtained by the Whipple collaboration in 2002 (continuous line) (Krennrich F. et al., 2002). 
This conclusion generalizes our result obtained with the analysis of the June 2008 flare (Aielli G. et al., 2010).

We investigated also the correlation between RXTE/ASM X-ray and ARGO-YBJ TeV fluxes (Fig. \ref{fig:mrk421-quadratic}). A positive correlation is clearly observed and the X-ray/TeV relation seems to be quadratic rather than linear, in agreement with the results of Fossati G. et al. (1998).

MGRO J1908+06 is a Pulsar Wind Nebula discovered by Milagro at a median energy of $\sim$20 TeV. Later, the Cherenkov telescopes HESS and VERITAS confirmed this observation. The Milagro and HESS energy spectra are in disagreement (Smith A.J. et al. 2009) probably due to the intrinsic extension of the source.
ARGO-YBJ observed a TeV emission from MGRO J1908+06 with a significance of 6 s.d. in about 800 days. We estimated its intrinsic extension in 0.48$\pm$0.28 deg, in agreement with the HESS determination. With this information we calculated a preliminary energy spectrum: dN/dE = (3.6$\pm$0.8)$\cdot$10$^{-13}\cdot$(E/6 TeV)$^{-2.2\pm 0.3}$ cm$^{-2}$ s$^{-1}$ TeV$^{-1}$ in the energy range 1-30 TeV. This spectrum is consistent with that measured by Milagro, in particular around 10 TeV, but is not consistent with the HESS calculation. The ARGO-YBJ data are not yet sensitive to the presence of an energy cutoff.

\section{Cosmic Ray Physics}

\subsection{Medium and Large scale anisotropies}

\begin{figure}[t!]
\begin{center}
\epsfysize=5.8cm \hspace{0.5cm}
\epsfbox{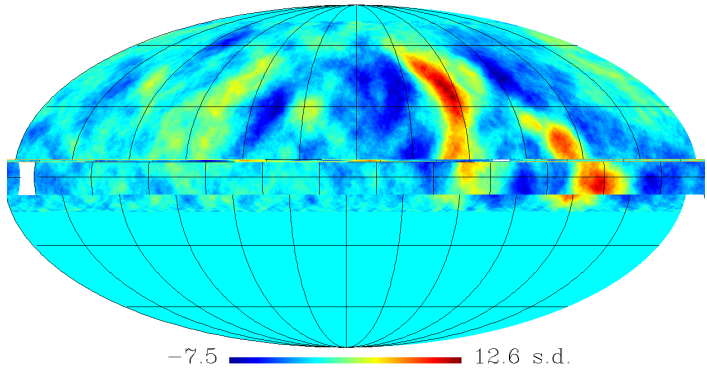} \vspace{-0.5cm}
\caption[h]{Medium scale anisotropy of CRs at energies $\sim$2 TeV. The color scale gives the statistical significance in standard deviations.} 
\label{fig:medium_anisot}
  \end{center}
\end{figure}
%
\begin{figure}[t!]
\begin{center}
\epsfysize=5.8cm \hspace{0.5cm}
\epsfbox{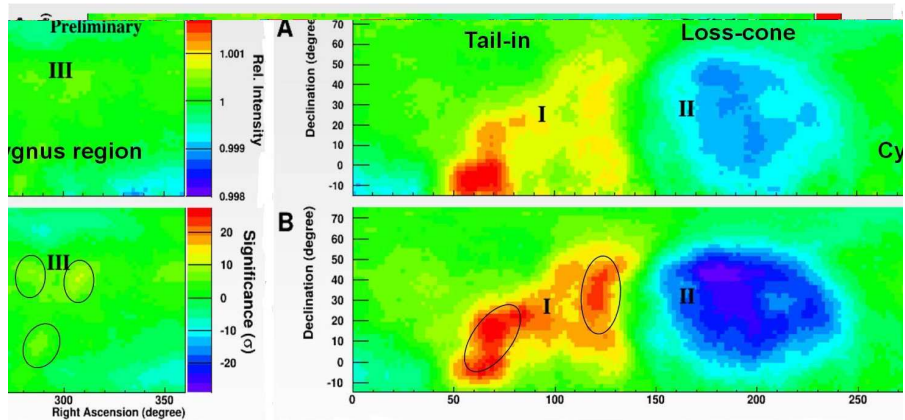} \vspace{-0.5cm}
\caption[h]{Large scale CR anisotropy observed by ARGO-YBJ at energies $\sim$2 TeV. In the upper plot the color scale gives the relative CR intensity, in the lower plot the statistical significance in standard deviations.} 
\label{fig:large_anisot}
  \end{center}
\end{figure}
%
Fig. \ref{fig:medium_anisot} shows the sky map for events with N$>$40 and zenith angle $\theta<40^{\circ}$ (corresponding to a proton median energy $\sim$2 TeV), obtained with a 5$^{\circ}$ smoothing radius. The analysis has been performed in order to be insensitive to the CR large scale anisotropy which is roughly one order of magnitude greater.
The map clearly shows two large hot spots in the region of the Galactic anticenter. The two excesses ($>$10 s.d., corresponding to a flux increase of $\sim$0.1\%) are observed by ARGO-YBJ around the positions $\alpha\sim$ 120$^{\circ}$, $\delta\sim$ 40$^{\circ}$ and $\alpha\sim$ 60$^{\circ}$, $\delta\sim$ -5$^{\circ}$, in agreement with a similar detection reported by the Milagro Collaboration (Abdo A.A. et al., 2008). However, the maximum of the second excess is slightly shifted towards lower declinations, probably because ARGO-YBJ can observe the southern regions of the sky with more efficiency, being located at a lower latitude than Milagro.
The deficit regions parallel to the excesses are due to a known effect of the analysis, that uses also the excess events to evaluate the background, artificially increasing the background.
The origin of this medium-scale anisotropy is puzzling. In fact, these regions have been interpreted as excesses of hadronic CRs, but TeV CRs are expected to be randomized by the magnetic fields. Understanding these anisotropies should be a high priority as they are probably due to a nearby source of CRs, as suggested by some authors (e.g., Markov M.A. et al., 2010).

With 2 years of data we carried out a 2D measurement to investigate the detailed structural information of the large scale CR anisotropy beyond the simple Right Ascension (R.A.) profiles. The so-called \textit{`tail-in'} and \textit{`loss-cone'} regions, correlated to an enhancement or deficit of CRs, are clearly visible with a significance of about 20 s.d. (Fig. \ref{fig:large_anisot}). The study of the large scale anisotropy is a useful tool in probing the magnetic field structure in our interstellar neighborhood as well as the distribution of sources.
The medium-scale anisotropy of Fig. \ref{fig:medium_anisot} is contained in the two hot spots in the large-scale anisotropy tail-in region.
A new excess component with a $\sim$0.1\% increase of the CR intensity in the Cygnus region is observed with a significance of about 10 s.d..
To quantify the scale of the anisotropy we fitted the 1D R.A. projection of the 2D map with the first two harmonics for three different energies. The preliminary results (E = 0.7 TeV, A$_1$=(3.6$\pm$0.1)$\cdot$10$^{-4}$, $\phi_1$=63.4$^{\circ}\pm$0.9; E = 1.5 TeV, A$_1$=(6.8$\pm$0.1)$\cdot$10$^{-4}$, $\phi_1$=41.0$^{\circ}\pm$0.7;
E = 3.9 TeV, A$_1$=(9.0$\pm$0.1)$\cdot$10$^{-4}$, $\phi_1$=35.3$^{\circ}\pm$0.6) are in good agreement with other experiments (Guillian G. et al., 2007).

\subsection{Measurement of the light component spectrum of CRs}

\begin{figure}[t!]
\begin{minipage}[t]{.47\linewidth}
\begin{center}
\epsfysize=4.8cm \hspace{0.5cm}
\epsfbox{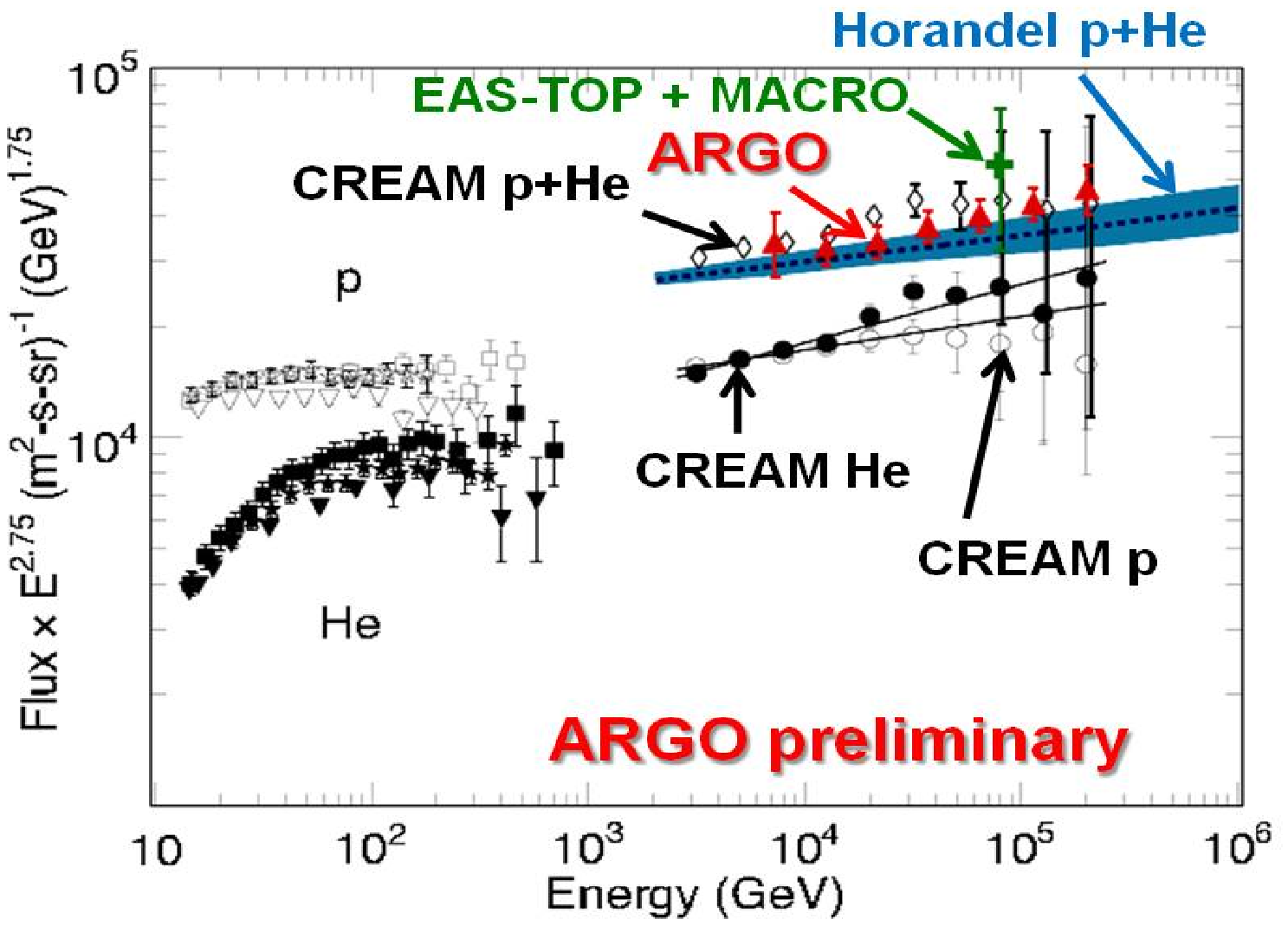} \vspace{-0.5cm}
\caption[h]{Light-component (p+He) spectrum of primary CRs measured by ARGO-YBJ compared with other experimental results.} 
\label{fig:light_spectrum}
  \end{center}
\end{minipage}\hfill
\begin{minipage}[t]{.47\linewidth}
  \begin{center}
\epsfysize=4.8cm \hspace{0.5cm}
\epsfbox{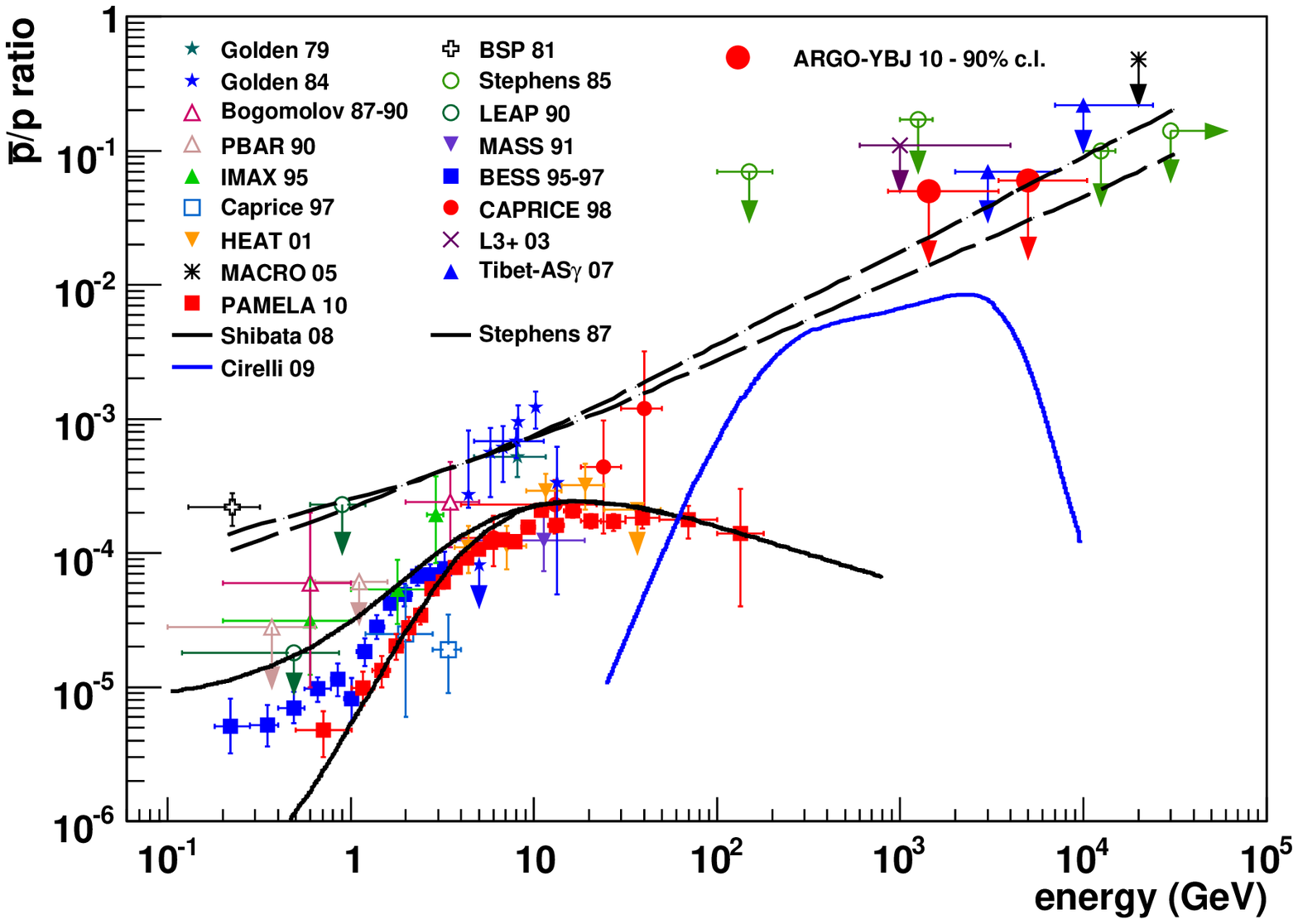}
\vspace{-0.5cm}
\caption[h]{The $\overline{p}/p$ ratio obtained with the
ARGO-YBJ experiment compared with all the available measurements and some theoretical models.} 
\label{fig:antip}
  \end{center}
\end{minipage}\hfill
\end{figure}
%

The differential energy spectrum of the primary CR light component (p+He) measured by ARGO-YBJ (filled triangles) in the energy region (5 - 250) TeV by using a Bayesian unfolding approach is compared with other experiments in Fig. \ref{fig:light_spectrum}.  
We selected showers with zenith angle $\theta<$30$^{\circ}$ and with reconstructed core position inside a fiducial area 50$\times$50 m$^2$ large by applying a selection criterion based on the particle density. In the energy range investigated the contamination of heavier nuclei is found to be negligible, not exceeding a few percent.
The ARGO-YBJ data agree remarkably well with the values obtained by adding up the proton and helium fluxes measured by CREAM both concerning the total intensities and the spectrum (Ahn H.S. et al., 2010). The value of the spectral index of the power-law fit representing the ARGO-YBJ data is -2.61$\pm$0.04, which should be compared to $\gamma_p$ = -2.66$\pm$0.02 and $\gamma_{He}$ = -2.58$\pm$0.02 obtained by CREAM.
The present analysis does not allow the determination of the individual proton and Helium contribution to the measured flux, but the ARGO-YBJ data are mainly induced by protons since the average energy of Helium primaries contributing to events with a given multiplicity is about 1.5 - 2 times greater than the average proton energy.
This result suggests a possible hardening of the proton spectrum at energies $>$5 TeV with respect to that obtained at lower energies from direct measurements with satellites and balloon-borne detectors.
We emphasize that for the first time direct and ground-based measurements overlap for a wide energy range thus making possible the cross-calibration of the different experimental techniques.

\subsection{Measurement of the $\overline{p}/p$ ratio at TeV energies}

In order to measure the $\overline{p}/p$ ratio at TeV energies we
exploit the Earth-Moon system as an ion spectrometer: if protons
are deflected towards East, antiprotons are deflected towards
West. If the energy is low enough and the angular resolution small
we can distinguish, in principle, between two shadows, one shifted
towards West due to the protons and the other shifted towards East
due to the antiprotons. If no event deficit is observed on the
antimatter side an upper limit on the antiproton content can be
calculated. 

We selected 2 multiplicity bins, 40$<$N$<$100 and N$>$100. In the first interval the statistical significance of the Moon shadow is 34 s.d., the measured angular resolution is $\sim$1$^{\circ}$ and the median energy is 1.4 TeV. For N$>$100 the significance is 55 s.d., the angular resolution is $\sim$0.6$^{\circ}$ and the median energy is 5 TeV.
Taking into account that at these energies protons are about 70\% of CRs, with all the data up to December 2009 we set two upper limits to the $\overline{p}/p$ ratio at the 90\% confidence level: 5\% at 1.4 TeV and 6\% at 5 TeV.
Since the spectral index of antiprotons around the energy range of interest is unknown, there is no reason to assume the proton spectral index.
Many unknown factors contribute to its value, mostly related to the diffusion coefficient inside the Galaxy. 
To investigate this point, different spectral indices were considered for primary antiprotons, $\gamma$= 2.0, 2.2, 2.4, 2.6, 2.8, 3.0. The limits of the antiproton/proton ratio vary of 20\% - 30\% with the spectral index, suggesting further studies on the possibility of setting some constraints on the diffusion coefficient of the CR propagation model.

The upper limits calculated with ARGO-YBJ are compared in Fig. \ref{fig:antip} with all the available $\overline{p}/p$ measurements and with some theoretical models for antiprotons production. The solid curves refer to a direct production model. The dashed lines refer to a model of primary $\bar{p}$ production by antigalaxies (Stecker F.W. and Wolfendale A.W., 1985). The rigidity-dependent confinement of cosmic rays in the Galaxy is assumed to be $\propto$ R$^{-\delta}$, and the two dashed curves correspond to the cases of $\delta$ = 0.6, 0.7. The dotted line refers to the contribution from a heavy Dark Matter particle annihilation (Cirelli M. et al., 2009).
We note that in the few-TeV range the ARGO-YBJ results are the lowest available, useful to constrain models for antiproton production in antimatter domains.

\section{Conclusions}

The ARGO-YBJ experiment is in stable data taking since November 2007 with a duty cycle $>$85\% and with excellent performance.
We observed the CR Moon and Sun shadows with a significance of 55 s.d. and 45 s.d., respectively. The detector performance is continuously monitored with the Moon shadow technique. The energy scale error is estimated less than 18\% and the overall absolute pointing error less than 0.1$^{\circ}$. The angular resolution has been stable at a level better than 10\% over the last 2 years.

Several interesting results are already available in cosmic ray physics as well as in gamma ray astronomy with the first 2 years of data.

We observed TeV emission from 3 sources with a significance greater than 5 s.d.: Crab Nebula, Mrk421 and MGRO J1908+06.
In particular we observed the Crab Nebula with a significance of 14 s.d. in $\sim$800 days.
A detailed long-term monitoring of the Mrk421 flaring activity has been carried out in the last 2 years. A clear correlation with X-ray data was found. The relation between spectral index and flux has been studied in one single flare (June 2008) and averaging the data over 2 years.


A medium-scale anisotropy has been observed with a significance greater than 10 s.d. at proton median energy of about 2 TeV. 
With 2 years of data we carried out a 2D measurement of the CR large-scale anisotropy to investigate detailed structural information beyond the simple Right Ascension profiles. We found a good agreement with other experiments.


The light-component (p+He) spectrum of primary CRs has been measured in the range 5 - 250 TeV. The preliminary results are in good agreement with the CREAM balloon data. For the first time direct and ground-based measurements overlap for a wide energy range thus making possible the cross-calibration of the different experimental techniques.

A measurement of the $\overline{p}/p$ ratio at few-TeV energies has been performed setting two upper limits at the 90\% confidence level: 5\% at 1.4 TeV and 6\% at 5 TeV.
In the few-TeV range these results are the lowest available, useful to constrain models for antiproton production in antimatter domains.

\bigskip
\bigskip
\noindent {\bf DISCUSSION}

\bigskip
\noindent {\bf JOSEF M. PAREDES:} Fermi and AGILE detected Cygnus X-3 recently. Did ARGO-YBJ detect any hint of signal from Cygnus X-3 at the epoch of the Agile and Fermi detections ?

\bigskip
\noindent {\bf G. DI SCIASCIO:} The monitoring of Cygnus X-3 is under way on different time scales. So far no clear detection at 5 standard deviation level has been observed in coincidence with satellite observations.

\bigskip
\noindent {\bf A. ERLYKIN:} How do you distinguish showers from light nuclei from those from heavier nuclei ?

\bigskip
\noindent {\bf G. DI SCIASCIO:} In the investigated energy range (5 - 250 TeV) the contribution of nuclei heavier than Helium to the ARGO-YBJ trigger is estimated less than 10\%. After selecting internal events by applying a criterion based on the particle density around the shower core, the contamination of heavier nuclei is found not exceeding a few percent.

\bigskip
\noindent {\bf T. MONTARULI:} What is your sensitivity to the Crab Nebula ?

\bigskip
\noindent {\bf G. DI SCIASCIO:} We observed the Crab Nebula with a significance of 14 s.d. in about 800 days, that is with a sensitivity of $\sim$50\% of the Crab flux per year, without any internal event selection or $\gamma$/hadron discrimination technique.

\end{document}